\documentclass[a4paper]{article}
\usepackage{fullpage}

\usepackage{graphicx}
\usepackage{hyperref}
\usepackage{dcolumn}
\usepackage{bm}

\usepackage{amsmath}
\usepackage{amssymb}
\usepackage{amsthm}

\usepackage{cite}
\usepackage{authblk}
\usepackage{comment}
\usepackage{enumerate}

\title{Critical fluctuations of the proton density in A+A collisions at $158A$ GeV}
\date{}

\author[21]{T.~Anticic}
\author[8]{B.~Baatar}
\author[6]{J.~Bartke}
\author[9]{H.~Beck}
\author[10]{L.~Betev}
\author[18]{H.~Bia{\l}\-{k}owska}
\author[9]{C.~Blume}
\author[20]{M.~Bogusz}
\author[18]{B.~Boimska}
\author[9]{J.~Book}
\author[1]{M.~Botje}
\author[10]{P.~Bun\v{c}i\'{c}}
\author[20]{T.~Cetner}
\author[1]{P.~Christakoglou}
\author[14]{O.~Chvala}
\author[15]{J.~Cramer}
\author[13]{V.~Eckardt}
\author[4]{Z.~Fodor}
\author[7]{P.~Foka}
\author[7]{V.~Friese}
\author[9,11]{M.~Ga\'{z}dzicki}
\author[20]{K.~Grebieszkow}
\author[7]{C.~H\"{o}hne}
\author[21]{K.~Kadija}
\author[10]{A.~Karev}
\author[8]{V.~I.~Kolesnikov}
\author[6]{M.~Kowalski}
\author[7]{D.~Kresan}
\author[4]{A.~Laszlo}
\author[1]{M.~van~Leeuwen}
\author[20]{M.~Ma\'{c}kowiak-Paw{\l}owska}
\author[17]{M.~Makariev}
\author[8]{A.~I.~Malakhov}
\author[16]{M.~Mateev}
\author[8]{G.~L.~Melkumov}
\author[9]{M.~Mitrovski}
\author[11]{St.~Mr\'owczy\'{n}ski}
\author[4]{G.~P\'{a}lla}
\author[2]{A.~D.~Panagiotou}
\author[20]{W.~Peryt}
\author[20]{J.~Pluta}
\author[15]{D.~Prindle}
\author[12]{F.~P\"{u}hlhofer}
\author[9]{R.~Renfordt}
\author[5]{C.~Roland}
\author[5]{G.~Roland}
\author[9]{A.~Rustamov}
\author[11]{M.~Rybczy\'{n}ski}
\author[6]{A.~Rybicki}
\author[7]{A.~Sandoval}
\author[13]{N.~Schmitz}
\author[9]{T.~Schuster}
\author[13]{P.~Seyboth}
\author[4]{F.~Sikl\'{e}r}
\author[19]{E.~Skrzypczak}
\author[20]{M.~Slodkowski}
\author[11]{G.~Stefanek}
\author[9]{R.~Stock}
\author[9]{H.~Str\"{o}bele}
\author[21]{T.~Susa}
\author[20]{M.~Szuba}
\author[3]{D.~Varga}
\author[2]{M.~Vassiliou}
\author[4]{G.~I.~Veres}
\author[4]{G.~Vesztergombi}
\author[7]{D.~Vrani\'{c}}
\author[11]{Z.~W{\l}odarczyk}
\author[11]{A.~Wojtaszek-Szwar\'{c}}
\author[ ]{\\(NA49 Collaboration)}
\affil[1]{NIKHEF, Amsterdam, Netherlands}
\affil[2]{Department of Physics, University of Athens, Athens, Greece}
\affil[3]{E\"{o}tv\"{o}s Lor\'{a}nt University, Budapest, Hungary}
\affil[4]{Wigner Research Center for Physics, Hungarian Academy of Sciences,
             Budapest, Hungary}
\affil[5]{MIT, Cambridge, Massachusetts, USA}
\affil[6]{H.~Niewodnicza\'{n}ski Institute of Nuclear Physics,
             Polish Academy of Science, Cracow, Poland}
\affil[7]{GSI Helmholtzzentrum f\"{u}r Schwerionenforschung,
             Darmstadt, Germany}
\affil[8]{Joint Institute for Nuclear Research, Dubna, Russia}
\affil[9]{Fachbereich Physik der Universit\"{a}t, Frankfurt, Germany}
\affil[10]{CERN, Geneva, Switzerland}
\affil[11]{Institute of Physics, Jan Kochanowski University, Kielce, Poland}
\affil[12]{Fachbereich Physik der Universit\"{a}t, Marburg, Germany}
\affil[13]{Max-Planck-Institut f\"{u}r Physik, Munich, Germany}
\affil[14]{Institute of Particle and Nuclear Physics, Charles
             University, Prague, Czech Republic}
\affil[15]{Nuclear Physics Laboratory, University of Washington,
             Seattle, Washington, USA}
\affil[16]{Atomic Physics Department, Sofia Univ. St.~Kliment
             Ohridski, Sofia, Bulgaria}
\affil[17]{Institute for Nuclear Research and Nuclear Energy, BAS, Sofia, Bulgaria}
\affil[18]{National Center for Nuclear Research, Warsaw, Poland}
\affil[19]{Institute for Experimental Physics, University of Warsaw,
             Warsaw, Poland}
\affil[20]{Faculty of Physics, Warsaw University of Technology, Warsaw, Poland}
\affil[21]{Rudjer Boskovic Institute, Zagreb, Croatia}

\author[2]{\\~\\N.~G.~Antoniou}
\author[2]{N.~Davis}
\author[2]{F.~K.~Diakonos}

\begin{document}

\maketitle
\thispagestyle{empty}

\hrule
\begin{abstract}
We look for fluctuations expected for the QCD critical point using an intermittency analysis in the transverse momentum phase space of protons produced around midrapidity in the 12.5\% most central C+C, Si+Si and Pb+Pb collisions at the maximum SPS energy of 158$A$~GeV. We  find  evidence  of  power-law  fluctuations  for  the  Si+Si  data. The fitted   power-law   exponent  $\phi_{2} = 0.96^{+0.38}_{-0.25}\text{ (stat.)}$ $\pm 0.16\text{ (syst.)}$ is consistent with the va\-lue expected for critical fluctuations. Power-law fluctuations had previously also been observed in low-mass $\pi^+ \pi^-$ pairs in the same Si+Si collisions.

\end{abstract}
\hrule


\newpage
\section{Introduction}
\label{sect1}

Theoretical investigations of the phase diagram of stro\-ngly interacting matter suggest the existence of a critical point (CP) at finite baryochemical potential and temperature. The prevailing view is that the CP corresponds to a second order phase transition which is considered as the endpoint of a line of first order transitions associated with the partial restoration of chiral symmetry when the temperature $T$, for given baryochemical potential $\mu_{B}$, increases beyond a critical value $T_c$ (for a review see \cite{Fukushima2011}). This hypothesis is compatible with results of Lattice QCD calculations \cite{Lattice1,Lattice2} although a generally accepted accurate prediction concerning the existence and location of the CP is not yet available. In current ion collision experiments at the SPS \cite{na49,na61} and RHIC \cite{RHIC1} an exploration of the QCD phase diagram is attempted by changing the energy and size of colliding nuclei. One of the main goals of this scanning program is to find evidence for the CP as a maximum of fluctuations in analogy to the phenomenon of critical opalescence in conventional matter \cite{Antoniou2006}. An experimental estimate of $(T,\mu_B)$ of the freeze-out state formed in the collisions is usually obtained from the observed particle yields \cite{Stachel2007,Becattini2006,Becattini2006b}.

A prerequisite for the experimental detection of the CP is to find suitable observables, as attempted in several recent theoretical studies \cite{Global,Local,Antoniou2005,Antoniou2006}. The order parameter of the phase transition is the chiral condensate $\langle \bar{q}q\rangle$ ($q$ is the quark field). The quantum state carrying the quantum numbers as well as the critical properties of the chiral condensate is the isoscalar $\sigma$-field. Assuming that this state can be formed in ion collisions there are two possibilities for its detection:

\begin{itemize}

\item Directly from its decay products~\cite{Antoniou2005}. The condensate, being unstable against changes of thermodynamic conditions (freeze-out), will decay mainly into pions at time scales characteristic of the strong interaction. The critical properties of the condensate are transferred to detectable 
$\pi^+ \pi^-$ pairs with invariant mass just above twice the pion mass \cite{Antoniou2005}. For an analysis of the expected fluctuations it is necessary to extract the pion pairs (dipions), which possess the critical correlations. This requires  removal  of  a  large  combinatorial background \mbox{\cite{Antoniou2005,NA49_PRC}}.

\item Through the mixing of the net-baryon density with the chiral condensate in a finite-density medium. The critical fluctuations are transferred to the net-baryon density \cite{Fukushima2011,Baryons,Antoniou2010}, which is an equivalent order parameter of the second order phase transition at the CP, as well as the net-proton density and the proton and antiproton densities separately. This is due to the direct coupling of the protons with the isospin zero $\sigma$-field~\cite{Hatta2003}. In contrast to the dipion case the critical correlations are carried by the observed protons and no reconstruction is needed. Thus detecting the QCD CP through fluctuations of the proton density is a very promising strategy. 
\end{itemize}

The experimental observables proposed for the search of the CP can be classified into two categories:

\begin{itemize} 
\item Event-by-event fluctuations of integrated quantities like multiplicity (variance \cite{NA49_n}, skewness and kurtosis \cite{STAR}) and averaged transverse momentum \cite{NA49_pt}. A maximum of these fluctuation measures as a function of size and energy of the colliding ions is expected in the vicinity of the CP. Indications of such a maximum were observed in Si+Si collisions at 158$A$ GeV~\cite{NA49_cp}.

\item Local power-law fluctuations \cite{Local} directly related to the critical  behaviour of density-density correlations. These have been shown to be detectable~\cite{Antoniou2005,Antoniou2006} in momentum space within the framework of an intermittency analysis \cite{Bialas1986} through the measurement of the scaled factorial moments (SF\-Ms) defined as: 
\begin{equation}
F_q(M)=\frac{ \langle \displaystyle{\frac{1}{M^D}\sum_{i=1}^{M^D}} n_i(n_i-1)..(n_i-q+1) \rangle }
{\langle \displaystyle{\frac{1}{M^D}\sum_{i=1}^{M^D}} n_i \rangle^q }
\label{eq:facmom}
\end{equation}
where $M^D$ is the number of equally sized cells in which the D-dimensional (embedding) space is partitioned, $n_i$ is the proton multiplicity in the i-th cell, and $q$ is the order (rank) of the moments. 
At the CP, the fluctuations of the order parameter, being self-similar \cite{Vicsek}, are described by a mono-fractal geometry \cite{Falconer} reflected  in  the  behavior  $F_q(M) \sim \left(M^D\right)^{\phi_q}$ of  the  SFMs  for  $M \gg 1$. The associated intermittency indices 
$\phi_q$ are predicted \cite{deWolf1996} to follow the pattern:
\begin{equation}
D \cdot \phi_q = (q-1) d_q ,
\label{eq:fdim}
\end{equation}
with $d_q$ the so called anomalous fractal dimension of the set formed by the order parameter density fluctuations. 
For a mono-fractal set $d_q$ is independent of $q$ and is related to the corresponding fractal dimension $d_F$ through the relation \cite{deWolf1996} $d_q=D-d_F$ . Based on universality class arguments, the exponents of the expected power-laws in transverse momentum space ($D=2$) were predicted to have the value $\phi_{2,cr}=\frac{2}{3}$ in case of the sigma condensate \cite{Antoniou2005} and $\phi_{2,cr}=\frac{5}{6}$ for the net-baryon density \cite{Antoniou2006}. Since the system we investigate is finite, we expect the power-law behaviour of the SFMs to hold only between two scales, dictated by the size of the system and the minimum distance of two protons in transverse momentum space \cite{Antoniou1998}. Moreover, in the course of intermittency analysis of experimental data, it was found necessary to remove a large background, (discussed in section \ref{sect2}) in order to reveal the predicted power-law exponents. 

In our analysis, due to the limited statistics, we will only consider the second scaled factorial  moment (SSFM) in transverse momentum space obtained by setting $q=2$ and $D=2$ in Eq.~(\ref{eq:facmom}). 
Much larger statistics than available is required in order to perform an intermittency analysis of higher than second order and to test the $q$ independence of the anomalous fractal dimensions $d_q$. 
This is due to the fact that analyzing  quantities related to the order parameter of the transition in the appropriate phase space region restricts significantly the mean multiplicity of the particles carrying the critical fluctuations (protons, $\pi^+$-$\pi^-$ pairs close to their production threshold). We use horizontal moments, which means that in Eq.(\ref{eq:facmom}) both the numerator and the denominator are separately averaged first over cells and then over events, for a given $M$ value. Thus, the denominator depends trivially on $M$, as $M^{-D \cdot q} \sim M^{-4}$, playing the role of normalization factor, with all non-trivial scaling lying in the numerator.

The intermittency analysis reported in this paper differs from previous studies~\cite{deWolf1996,KLM,Hwa1991} in measuring quantities associated with the order parameter of the phase transition.
In Ref.~\cite{KLM} all charged particles in the entire available phase space were used, allowing also higher moment calculations. However, in this case the quantitative relation with the predicted power law is lost.
\end{itemize}

In the present work we report on intermittency analysis of protons produced from A+A collisions in the NA49 experiment at the CERN SPS at beam energy of 158$A$ GeV. This analysis is a continuation of our search for indications of the CP in local density fluctuations of low-mass $\pi^+ \pi^-$ pairs~\cite{NA49_PRC}.

\section{Intermittency in the presence of background}
\label{sect2}

Following the methodology of Refs.~\cite{Antoniou2005,Antoniou2006,Bialas1986} and using Eq.~(\ref{eq:facmom}) for $q=2$ and $D=2$, we calculate for every considered ensemble of protons the SSFM as a function of the number $M$ of subdivisions in each transverse momentum space direction of a rectangular domain $\mathcal{D}$. In Eq.~(\ref{eq:facmom}) $n_i$ is the number of protons in the cell $i$ and $M^2$ is the total number of cells. The brackets $\langle ... \rangle$ indicate averaging over events, however in the following, in order to simplify the notation, we will use $\langle ... \rangle$ to indicate averaging over both, bins and events and drop the subscript $i$ referring to individual cells.  
For a pure critical system freezing out exactly at the chiral CP the SSFM (Eq.~(\ref{eq:facmom})) of protons emitted into a small window around midrapidity is expected to possess a \mbox{2-D} power-law dependence $F_2(M) \sim M^{2 \phi_{2,cr}}$ for $M \gg 1$, with exponent (intermittency index) $\phi_{2,cr}=\frac{5}{6}$ determined by universality class arguments associated with the critical properties of QCD \cite{Antoniou2006}. This results from the fact that the density of the critical system is approximately constant in center of mass rapidity $y$ whenever $\sinh y \approx y$, which is valid at midrapidity (within a 10\% approximation for $|y| \leq 0.8$). Then the fluctuations in transverse momentum and rapidity space factorize, becoming statistically independent \cite{Antoniou2006}. 

In practice the critical system will never be 100\% pure or complete, for a number of reasons. First of all, there will always be some non-proton tracks that are misidentified as protons, as well as true proton tracks erroneously rejected. Moreover, due to the corona effect only part of the system may reach the deconfined phase and the freeze-out may be located at a distance from the CP. Thus, the ensemble of protons will be contaminated by a percentage of non-critical protons which will deform $F_2(M)$, leading to a decrease of the value of $\phi_2$ and/or a modification of the power-law behaviour~\cite{Antoniou2005,NA49_PRC}. Assuming that the proton multiplicity in each cell can be divided into background and critical contributions, $n = n_b + n_c$ one can write the numerator in Eq.~(\ref{eq:facmom}) as:
\begin{equation}
\langle n(n-1) \rangle = \langle n_c(n_c-1) \rangle + \langle n_b(n_b-1) \rangle + 2 \langle n_b  n_c \rangle 
\label{eq:crit2p}
\end{equation}
where $\langle n_c(n_c-1) \rangle$ is the critical component in the phase space integrated two-particle density, $\langle n_b(n_b-1) \rangle$ is the background contribution due to the presence of non-critical protons and $2 \langle n_b n_c \rangle$ is a cross-term. The latter vanishes when $\langle n_c \rangle \to 0$ or $\langle n_b \rangle \to 0$. Under general conditions we can write the cross-term in Eq.~(\ref{eq:crit2p}) as $2 \langle n_b n_c \rangle =2 \langle n_b \rangle \langle n_c \rangle f_{bc}$ where $f_{bc}$ is  a finite quantity which cannot be further determined. Dividing both sides of Eq.~(\ref{eq:crit2p}) by $\langle n \rangle^2$ (which is proportional to $(M^{-2})^{2}$ for large $M$) we obtain the expression:
\begin{equation}
\Delta F_2(M) = F_2^{(d)}(M)-\lambda(M)^2 F_2^{(b)}(M)  -2\lambda(M) \left( 1 - \lambda(M) \right) f_{bc}
\label{eq:correlator}
\end{equation}
for the correlator $\Delta F_2(M) = \langle n_c(n_c-1) \rangle / \langle  n \rangle^2$ containing the critical contribution. In the right hand side of Eq.~(\ref{eq:correlator}) $F_2^{(d)}(M)$ is the SSFM calculated from the data, $\lambda(M)=\langle  n_b \rangle / \langle n \rangle$ is a measure of the contamination by non-critical protons and $F_2^{(b)}(M) = \langle n_b(n_b-1) \rangle / \langle n_b \rangle^2$ is the SSFM of the background. Note that for $M \gg 1$ the ratio $\lambda$ becomes independent of $M$ and can be identified as the fraction of non-critical protons in the considered ensemble. By construction the correlator $\Delta F_2(M)$ possesses the same $M^2$ dependence as the SSFM of the critical component ($\Delta F_2(M) \sim M^{2 \phi_{2,cr}}$) for $M \gg 1$ since the numerator carries the non-trivial scaling and background contribution to the two-particle correlations is removed.

Two special cases of Eq. (\ref{eq:correlator}) merit discussion:
\begin{enumerate}
\item When background dominates, as in the present analysis, $\lambda \lesssim 1$. In this case we neglect the third term in the rhs of Eq.~(\ref{eq:correlator}) which is equivalent to omitting the cross-term in Eq.~(\ref{eq:crit2p}). However, the difference of the first two terms in Eq.(\ref{eq:correlator}) can be comparably small, and therefore the justification of this approximation is non-trivial and will be provided by model simulations. These will be discussed in Section~\ref{sect4} together with the presentation of the results from the data analysis. 
    
\item When the freeze-out of the considered system occurs very close to the chiral CP, there is a possibility that the background contribution becomes very small so that $\lambda \gtrsim 0$ for $M \gg 1$ and the correlator $\Delta F_2(M)$ coincides with $F_2^{(d)}(M)$.
\end{enumerate}

In the first case one can use mixed events generated from the data to simulate the background contribution and estimate $F_2^{(b)}(M)$ in Eq.~(\ref{eq:correlator}), assuming that the background consists of particles uncorrelated in transverse momentum space. By construction, $\langle n \rangle_{mixed} = \langle n \rangle$ and  $F_2^{(m)}(M) \simeq F_2^{(b)}(M)$, since  we  use  scaled  factorial  moments.

\section{Data and methods of analysis}
\label{sect3}

The analysed data were recorded by the NA49 experiment~\cite{na49} in  A+A  collisions  at   maximum   CERN   SPS   energy  of \mbox{$158A$ GeV} ($\sqrt{s_{NN}} = 17.3$ GeV). For the analysis we used the most central collisions (12\%, 12\%, 10\%) of ``C", ``Si" and Pb nuclei on C (2.4\% interaction length), Si (4.4\%) and Pb (1\%) targets, respectively. The incident C and Si nuclei were produced by
fragmentation of a Pb beam of 158$A$~GeV~beam energy~\cite{na49} and were selected by 
magnetic rigidity ($Z/A = 0.5$) and by specific energy loss in multiwire proportional chambers in the beam line upstream of the target position.
The ``C" beam as defined by the online trigger and offline selection was a mixture of ions with charge $Z=6$ and $7$ (intensity ratio 69:31); the ``Si" beam of ions with $Z=13, 14$ and $15$ (intensity ratio 35:41:24)~\cite{NA49_kstar}. The trigger selected the centrality of the collisions based on a measurement of the energy deposi\-ted by projectile spectator nucleons in a downstream ca\-lorimeter.
The event statistics amounted to 148k events for ``C"+C, 166k events for ``Si"+Si, and 330k events for Pb+Pb. The standard event and track selection cuts of the NA49 experiment were applied as described in~Ref.\cite{NA49_pL2011}.

Specifically, to avoid double counting of split tracks we only accepted tracks for which the ratio of number of measured points to estimated maximum number of points in the TPCs exceeds 55\%. We made sure that the resulting sample of particle tracks was not contaminated by fake close-pairs which could affect the intermittency analysis by examining the two-track distance distribution and the two-particle correlation function in $q_{inv}$ (discussed below).

Proton identification~\cite{NA49_pL2011} used the measurements of particle energy loss $dE/dx$ in the gas of the time projection chambers. The inclusive $dE/dx$ distribution for positively charged particles in each reaction was fitted in 10 bands of momentum $p$ to a sum of contributions $f^{\alpha}(dE/dx,p)$ from different particle species $\alpha$ with $\alpha$ = $\pi$, $K$, $p$, $e$. The probability $P$ for a track with energy loss $x_{i}$ and momentum $p_{i}$ of being a proton is then given by $P= f^{\text{p}}(x_{i},p_{i}) / (f^{\pi}(x_{i},p_{i})+f^{\text{K}}(x_{i},p_{i}) + f^{\text{p}}(x_{i},p_{i})+f^{\text{e}}(x_{i},p_{i})) $. The value of $P$ for proton candidates had to exceed 80\% for the ``C"+C and ``Si"+Si systems and 90\% for Pb+Pb collisions. 

We calculated the SSFMs according to Eq.~(\ref{eq:facmom}) for the three considered systems ``C"+C, ``Si"+Si and Pb+Pb in the domain $\mathcal{D}=[-p_{x,max},p_{x,max}] \otimes [-p_{y,max},p_{y,max}]$ of the transverse momentum plane $(p_x,p_y)$ with $p_{x,max} = p_{y,max} =\- 1.5$ $\text{GeV/c}$. This plane is perpendicular to the beam direction, $p_x$, $p_y$ being the corresponding horizontal and vertical transverse momentum components. For the calculations we selected protons  with center of mass rapidity $\vert y_{CM} \vert \leq 0.75$ in order to restrict the analysis to the midrapidity region which is a necessary condition for the appearance of power-law critical fluctuations (see Section~\ref{sect2}). This selection also removes those protons which underwent only one diffractive interaction as is the case for nucleons in the corona. The mean proton multiplicities in the considered
rapidity-transverse momentum interval were: $\langle p \rangle_{\text{``C"+C}} = 1.6 \pm 0.8$, $\langle p \rangle_{\text{``Si"+Si}} = 3.1 \pm 1.7$ and $\langle p \rangle_{\text{Pb+Pb}} = 9.1 \pm 3.2$, respectively, where the second set of numbers give the standard deviation of the multiplicity distributions.

Usually in an intermittency analysis the only change of the transverse momentum lattice is the decrease of the lattice constant with increasing $M$ (the lattice becomes finer), keeping the lattice position in transverse momentum space fixed. However, the lattice bin boundaries may split proton pairs at distances smaller than the bin size, separating protons into different bins, which leads to unwanted bin-to-bin fluctuations. When the multiplicity per event is large these fluctuations are not relevant compared to the bin-to-bin fluctuations generated by the protons within a single bin. Since in our analysis, especially for the light systems (``C" and ``Si"), the multiplicity per event is small, this splitting effect due to the boundaries becomes important. To reduce these fluctuations, we calculated the SSFMs using slightly displaced transverse momentum lattices and averaged the resulting $F_2(M)$ values over the lattice positions. The displacement of the lattices was chosen small enough in order to avoid significant changes of the content of the analysed ensemble of protons. We used 11 differently located lattices shifting their position from $-1.53$ GeV/$c$ $ < p_{x,y} < 1.47$ GeV/$c$ to $-1.47$ GeV/$c$ $ < p_{x,y} < 1.53$ GeV/$c$. This procedure was applied to both the analysed data and the mixed events generated from the corresponding data.

Having calculated the SSFMs of the data $F_2^{(d)}(M)$ and the associated mixed events $F_2^{(m)}(M)$ as described above we subsequently estimated the correlator Eq.~(\ref{eq:correlator}) by the difference 
\begin{equation}
\Delta F_2^{(e)}(M) = F_2^{(d)}(M)  - F_2^{(m)}(M)
\label{eq:estimator}
\end{equation}
where we set $\lambda = 1$  and thus neglected the third term (cross-term) in the rhs of Eq.~(\ref{eq:correlator}). In section~\ref{sect4} we justify this choice by simulating and analysing data sets consisting of
a mixture of critical protons generated by the Critical Monte-Carlo (CMC) code \cite{Antoniou2006,Local} and randomly distributed protons. 

We searched for an intermittency effect $\Delta F_2^{(e)}(M) \sim$ \phantom{} $(M^2)^{\phi_2}$ for $M^2 \gg 1$ \cite{Bialas1986,Tannenbaum1995}. The intermittency index $\phi_2$ can be determined by a power-law fit (PF) to $\Delta F_2^{(e)}(M)$ in the region of sufficiently large $M^2$. Standard error propagation turned out to be inadequate for our intermittency analysis~\cite{Metzger}. We therefore obtained estimates of the statistical uncertainties of the SSFMs as well as of the $\phi_2$ values by using the resampling method~\cite{Efron1979}. This method involves constructing new sets of events out of the original one, containing the same number of events as the original. This is achieved by uniform and random sampling of events, with replacement, from the original set, so that a given event may be drawn any number of times. In the new sets, some events are of necessity omitted and others duplicated. We then calculate SSFMs as well as $\Delta F_2^{(e)}(M)$ for each resampled set~\cite{Metzger}.  In the calculation of $\Delta F_2^{(e)}(M)$ we used for all samples the same, very large (of the order of $10^7$) set of mixed events (one for each considered system), in order to reduce the statistical fluctuations of the background and the required computing time. Then we determined through a PF the intermittency index $\phi_2$ for each sample and obtained the distribution $P(\phi_2)$. 

\begin{figure}
\includegraphics[width=\columnwidth]{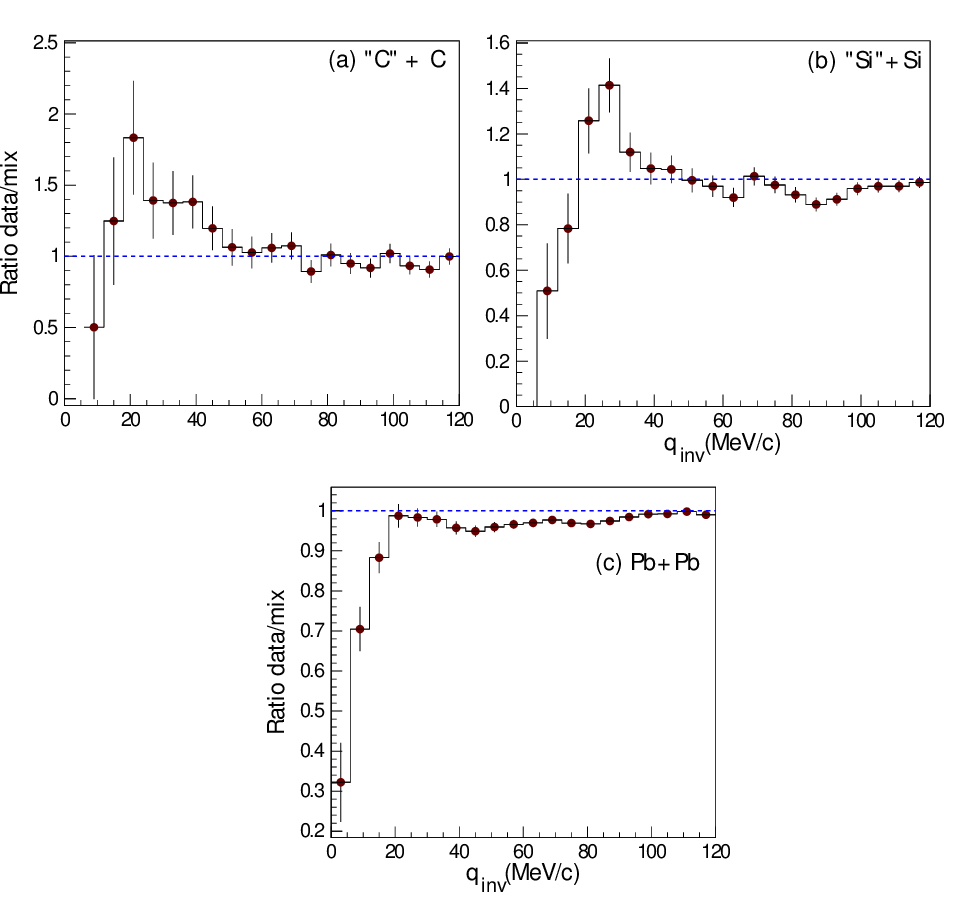}
\caption{(Color online) The $C(q_{inv})$ correlation function of proton pairs (ratio of pairs from real and mixed
events) at midrapidity ($-0.75 < y_{CM} <0.75$) for the most central collisions of (a) ``C"+C (centrality 12\%), (b) ``Si"+Si (centrality 12\%) and (c) Pb+Pb  (centrality 10\%) at $\sqrt{s_{NN}}=17.3$ GeV. }
\label{fig:qinv}
\end{figure}

A comment is now in order. Although the correlator\phantom{ } $\Delta F_2^{(e)}(M)$ takes care of the background of uncorrelated protons, proton-proton correlations due to Coulomb repulsion and Fermi-Dirac statistics, which have nothing to do with the CP, still remain and must be removed before performing intermittency analysis. To study the effect of these anti-correlations we calculated the distribution of the relative four-momenta of the proton pairs $q_{inv}=\frac{1}{2}\sqrt{-(p_1-p_2)^2}$ and the associated correlation function (ratio of true to mixed-event pairs) for all investigated systems (``C''+C, ``Si''+Si, Pb+Pb). The results of this calculation are shown in Fig.~\ref{fig:qinv}. As expected, the distribution develops a dip in the low $q_{inv}$ region due to Fermi-Dirac statistics and Coulomb repulsion followed by a strong-interaction maximum around $20$ MeV/c which, in agreement with theoretical predictions \cite{Koonin1977}, becomes more pronounced with decreasing size of the colliding nuclei. This behaviour suggests the introduction of a lower $q_{inv}$-cutoff in the selection of proton tracks for the intermittency analysis in order to remove the effect of these unwanted non-critical correlations. In addition we note that the absence of any peak in the limit $q_{inv} \to 0$ demonstrates the absence of split tracks in all three reactions which could compromise the intermittency analysis.

\begin{figure}
 \includegraphics[width=\columnwidth]{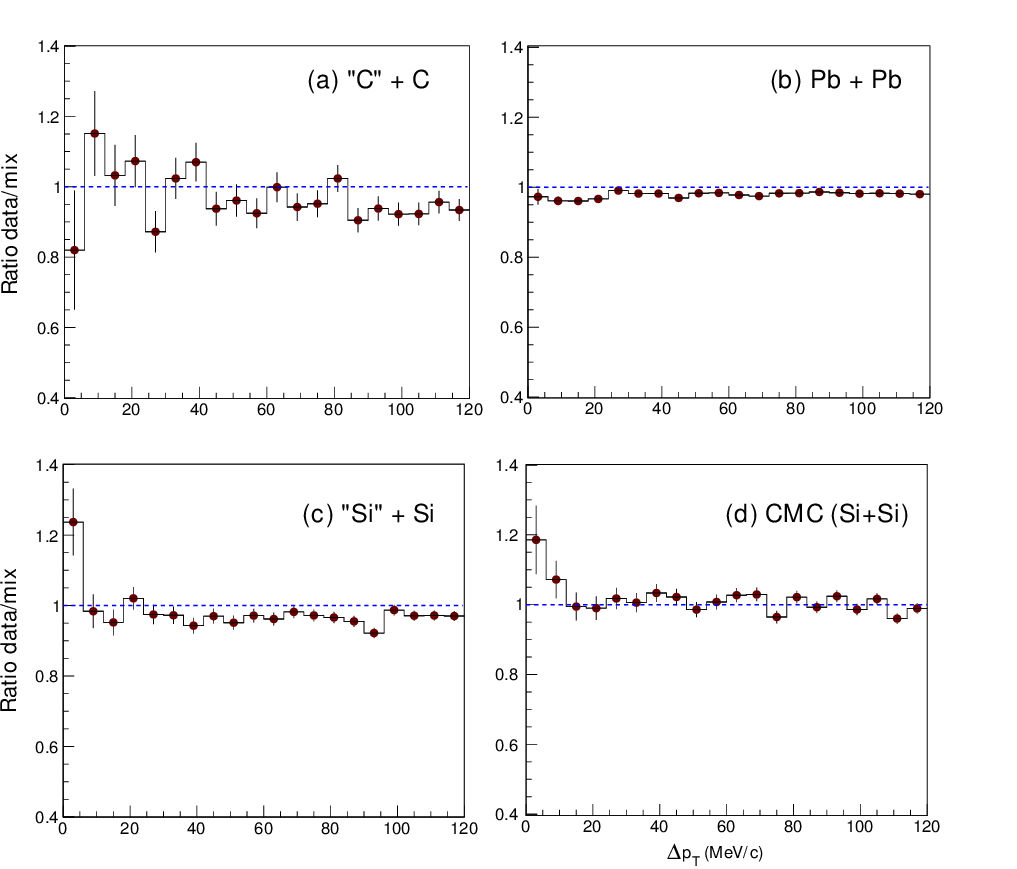}
 \caption{(Color online) The $C(\Delta p_T)$ correlation function of proton pairs (ratio of pairs from real and mixed events) at midrapidity ($-0.75 < y_{CM} <0.75$) for the most central collisions of (a) ``C"+C (centrality 12\%), (b) Pb+Pb  (centrality 10\%), (c) ``Si"+Si (centrality 12\%) at $\sqrt{s_{NN}}=17.3$ GeV, as well as for (d) CMC simulated Si+Si events (99\% noise).}
 \label{fig:DpT}
\end{figure}

Finally, since we study the correlations of protons in the transverse momentum space, we looked for evidence of strong correlations in the low relative $p_T$ region, i.e. for proton tracks that are close in transverse momentum space. To this end, we calculated the distribution in $\Delta p_T$ 
\begin{equation}
 \Delta p_T = \frac{1}{2}\sqrt{(p_{X1}-p_{X2})^2 + (p_{Y1}-p_{Y2})^2}
\end{equation}
the difference in $p_T$ of protons in the pairs, as well as the associated correlation function.
The results of the calculation are shown in Fig.~\ref{fig:DpT}, where we also plotted the $\Delta p_T$ distribution for a simulated CMC data set corresponding to a critical system mixed with 99\% random proton tracks. We see from Fig.~\ref{fig:DpT} that ``C''+C and Pb+Pb data sets do not exhibit significant correlations in the low $\Delta p_T$ region, whereas ``Si''+Si shows a peak at low $\Delta p_T$, which is comparable to the behaviour of the simulated CMC dataset.

\section{Results}
\label{sect4}

\begin{figure}
\includegraphics[width=\columnwidth]{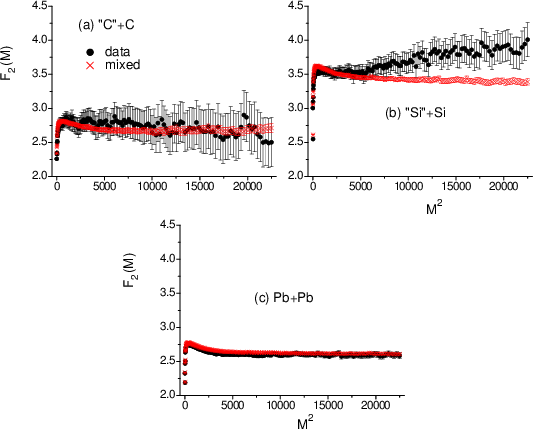}
\caption{(Color online) SSFMs of the proton density in transverse momentum space at midrapidity ($-0.75 < y_{CM} <0.75$) for the most central collisions  of (a) ``C"+C (12\%), (b) ``Si"+Si (12\%), and (c) Pb+Pb (10\%) at $\sqrt{s_{NN}}=17.3$ GeV. The circles (crosses)  represent the SSFM $F_2(M)$ of the data (mixed events) respectively. Error bars were obtained from the resampling method.}
\label{fig:F2M}
\end{figure}

The results for the SSFM $F_2(M)$ versus $M^2$ of the three analysed systems are shown with black circles in Fig.~\ref{fig:F2M}. In the same figure we also plot the SSFMs for the corresponding mixed events (red crosses). We used a universal cut $q_{inv} \geq 25$ MeV/c for all analysed proton pairs to take care of anti-correlations induced by Coulomb repulsion and Fermi-Dirac statistics, removing both proton tracks whenever this criterion was not fulfilled. For the ``Si"+Si system we observe that for large $M^2$ values the SSFMs of the data are clearly larger than those of the mixed events. The difference between the two moments increases with increasing number of cells $M^2$, a typical characteristic of intermittent behaviour. This observation is an indication for sizeable correlations among the produced protons. However, due to the small event ensemble, ``Si''+Si $F_2$ values are accompanied by large statistical errors, which are calculated as the resampling method variances of the lattice averaged $F_2$ values. In the ``C"+C and Pb+Pb cases the SSFM values of the data and the background overlap especially in the region of large $M^2$ values. This suggests the absence of an intermittency effect in these systems. The maximum number of bins used for the calculation of the SSFMs is $M_{max}=150$ per $p_T$ direction leading to a minimum bin size of $20$ MeV/$c$. The latter is much greater than the experimental resolution $\delta p_T$ in this phase space region ($\delta p_T \lesssim 5$ MeV/$c$). Thus systematic errors due to transverse momentum uncertainties are negligible for our analysis.

\begin{figure}
\includegraphics[width=\columnwidth]{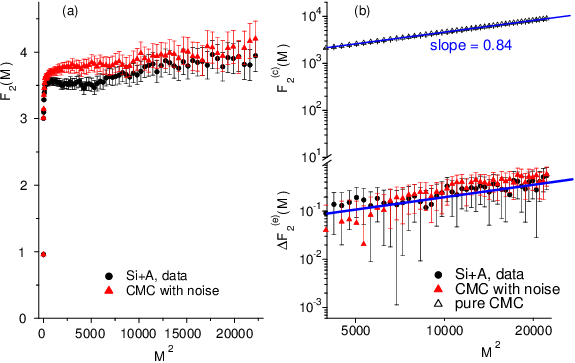}
\caption{(Color online) (a) SSFM of the proton density in transverse momentum space (filled triangles) for 150k events generated by the CMC code to simulate central collisions of the ``Si"+Si system at $\sqrt{s_{NN}}=17.3$ GeV. The critical system is contaminated with probability $\lambda=99\%$ with uncorrelated random tracks. For comparison we also show the corresponding result for the SSFM obtained from the ``Si"+Si data (filled circles), (b) The SSFM $F_2(M)$ of the 150k CMC events without contamination (open triangles) as well as the estimator  $\Delta F_2^{(e)}(M)$ for the contaminated ensemble (filled triangles) and the ``Si"+Si system (filled circles) in double logarithmic scale. Power-laws lines of slope $\phi_2 = 0.84$ are plotted as a visual guide. Only the region $M^2 > 1000$ is displayed in (b).}
\label{fig:F2MCMC}
\end{figure}

We used a proton generating modification of the CMC code~\cite{Antoniou2006,Local} to simulate our experimental results (see Fig.~\ref{fig:F2MCMC}) for the ``Si" + Si system employing a stochastic process generating a random fractal with dimension\footnote{The dimension $d_F$ refers to the geometry in transverse momentum space, denoted in \cite{Antoniou2006} by $\tilde{d}_F$. There is a corresponding fractal in transverse configuration space. The two fractal sets are related by a Fourier transform.} $d_F=\frac{1}{3}$, leading to an intermittency index of $\phi_2 = \frac{5}{6}$, which is exactly the value expected for the critical system according to Eq.(\ref{eq:fdim}). The SSFM in transverse momentum space of the pure critical system attains very large values for $M^2 \gg 1$  as shown in Fig.~\ref{fig:F2MCMC}~b. In order to reduce them to the level of values observed in Fig.~\ref{fig:F2M} one has to contaminate the critical system with a dominant random component. In practice we replaced  tracks of the critical ensemble by random tracks with probability $\lambda$. We illustrate this procedure in Fig.~\ref{fig:F2MCMC}. The random tracks were selected to respect the transverse momentum distribution of the ``Si"+Si data. The result of the simulation is shown by filled triangles in Fig.~\ref{fig:F2MCMC}a. For comparison we also plotted the SSFM of the corresponding data (filled circles, data points from Fig.~\ref{fig:F2M}~b). The fraction of random protons required by the data in the simulation turned out to be $\lambda=0.99~$ which is very close to one. Actually this $\lambda$ value coincides with the asymptotic value ($M^2 \gg 1$) appearing in Eq.~(\ref{eq:correlator}). Thus for the considered system the background contribution can be well simulated by mixed events justifying the use of Eq.~(\ref{eq:estimator}) to estimate the corresponding correlator. This is clearly illustrated in Fig.~\ref{fig:F2MCMC}b where we plot in log-log scale the SSFM of the pure critical system (open triangles) together with the estimator of the correlator $\Delta F_2^{(e)}(M)$ (filled triangles) determined by applying Eq.~(\ref{eq:estimator}) to the contaminated CMC data. It is evident that, although the two quantities differ by orders of magnitude, the exponent of the underlying power-law is the same, allowing the determination of the intermittency index $\phi_{2,cr}$ through the experimentally accessible $\Delta F_2^{(e)}(M)$. For comparison we include in Fig.~\ref{fig:F2MCMC}b also the estimated correlator $\Delta F_2^{(e)}(M)$ for the ``Si"+Si system (filled circles). Good agreement with the result of the contaminated CMC is observed also at the level of correlators. Note that the property of $\lambda$ being very close to one is robust, being determined by the order of magnitude of the SSFM values of the data for large $M^2$. According to the discussion in Section~\ref{sect2}, one possible scenario for the large background is that the freeze-out state of ``Si" lies at some distance from the CP within the critical region. In this case, if one succeeded in further approaching the CP by changing the energy and the size of the colliding ions, one would expect a rapid decrease of $\lambda$. However, hadronization and subsequent rescattering may mask this effect in experimental data.

\begin{figure}
 \includegraphics[width=\columnwidth]{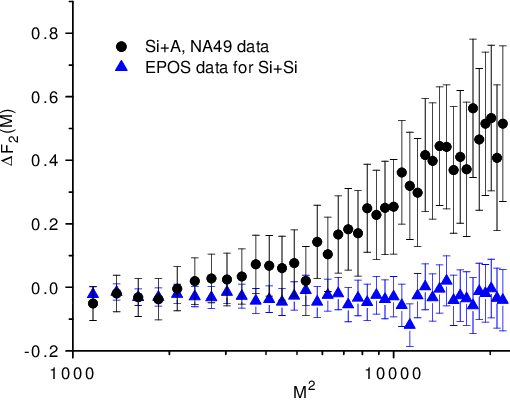}
 \caption{(Color online) The estimated correlator $\Delta F_2^{(e)}(M)$ of protons for the 12\% most central collisions (a) from the EPOS event generator (blue triangles), and (b) from ``Si"+Si data (black circles) at $\sqrt{s_{NN}}=17.3$ GeV. Error bars were obtained from the resampling method.}
 \label{fig:epos}
\end{figure}

We also checked our experimental results against simulated events produced by the EPOS event generator~\cite{EPOS} which includes high-$p_T$ jets, in order to examine whether the presence of a small number of protons in these jets can produce an intermittency effect. To this end, we configured EPOS to generate a set of 630k events corresponding to a beam of Si nuclei on a Si target (Z=14, A=28, for both beam and target), with a maximum impact parameter of b = 2.6 fm, corresponding to the centrality (12\%) of the ``Si''+Si experimental dataset. The center of mass energy was set at $\sqrt{s_{NN}}=17.3$ GeV, and $p_T$, $p_{tot}$ and rapidity cuts were applied exactly as in the NA49 data. Finally, we performed an intermittency analysis of protons in transverse momentum space for the simulated events, as well as the corresponding mixed events. Figure~\ref{fig:epos} compares the correlator $\Delta F_2^{(e)}(M)$ of the EPOS events with that from the ``Si''+Si data. It is evident that EPOS, including conventional sources of correlation, for example jet production and resonance decays, cannot account for the intermittency observed in ``Si''+Si, since its correlator fluctuates around or below zero.

\begin{figure}
\includegraphics[width=\columnwidth]{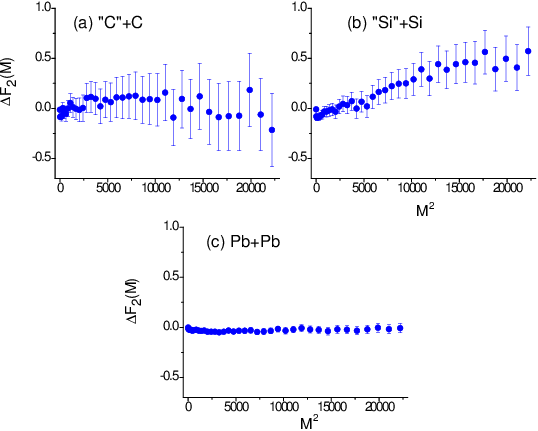}
\caption{(Color online) The estimated correlator $\Delta F_2^{(e)}(M)$ corresponding to the moments of Fig.~\ref{fig:F2M}, for the  most central collisions of (a) ``C"+C (centrality 12\%), (b) ``Si"+Si (centrality 12\%)  and (c) Pb+Pb (centrality 10\%) at $\sqrt{s_{NN}}=17.3$ GeV. Error bars were obtained from the resampling method.}
\label{fig:DF2M}
\end{figure}

The  panels (a),(b),(c) in Fig.~\ref{fig:DF2M} show the estimator of the correlator $\Delta F_2^{(e)}(M^2)$ as a function of $M^2$ for the original data sets ``C"+C, ``Si"+Si and Pb+Pb respectively. For ``Si"+Si an intermittency effect shows up for $M^2 > 6000$ (see Fig.~\ref{fig:F2M}b). Therefore we used this as the lower value of $M^2$ ($M^2_{min}$) in our fits. The intermittency index $\phi_2$ for the ``Si" + Si system was then determined from a PF to the corresponding correlator $\Delta F_2^{(e)}(M)$. For ``C"+C and Pb+Pb the values of $\Delta F_2^{(e)}(M^2)$ scatter
around zero. Therefore an intermittency effect is not present in these two systems.

\begin{figure}
\includegraphics[width=\columnwidth]{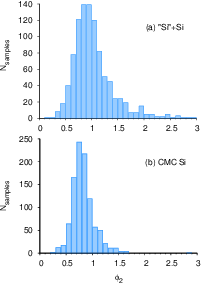}
\caption{(Color online) The distribution $P(\phi_2)$ obtained applying the resampling method to calculate $\phi_2$ for (a) protons produced in the 12.5\% most central collisions of  ``Si"+Si at $\sqrt{s_{NN}}=17.3$ GeV, and (b) protons generated by the CMC code ($\lambda$ = 99\% contamination level). $N_s$ = 1000 samples were produced for both cases.}
\label{fig:Pphi2}
\end{figure}

When applying the resampling method we constructed 1000 samples for ``Si" +Si and calculated for each sample the correlator $\Delta F_2^{(e)}$ as well as the corresponding $\phi_2$ value. The obtained distribution $P(\phi_2)$, shown in Fig.~\ref{fig:Pphi2}a, is highly asymmetric.  Using the resampling technique we also calculated the $P(\phi_2)$ distributions for a noise contaminated CMC data set with mean multiplicity equal to that of the ``Si"+Si system. In the simulation we can evaluate Eq.~(\ref{eq:correlator}) directly. When the cross-term in Eq.~(\ref{eq:correlator}) is taken into account, the corresponding $\phi_2$ distribution of the CMC model becomes a delta-like function centered at $0.8382(6)$. This is very close to the theoretically predicted value of $\frac{5}{6}$ which was used in the simulation. Omitting the cross-term results in a spread of the $\phi_2$ values around this central value. In fact using Eq.~(\ref{eq:estimator}) for the correlator of the contaminated CMC data set we find the estimated (median) $\phi_2$ value $0.80^{+0.19}_{-0.15}$ for the ``Si"+Si simulation. The corresponding distribution $P(\phi_2)$ is shown in Fig.~\ref{fig:Pphi2}b. The distance of the median from the expected value is much smaller than the spread of values, i.e. the median is almost unbiased. The use of the estimator (\ref{eq:estimator}) in our analysis therefore allows us to determine the intermittency index $\phi_2$ in a noise dominated data set.

Due to the skewness of $P(\phi_2)$, the appropriate estimator of $\phi_2$ is the median, and the measure of the associated statistical uncertainty is a confidence interval between two quantiles cutting off an equal lower and upper percentage of the distribution \cite{Statbook}. We chose the sextiles (partitioning the distribution into sixths), which correspond to a 67\% confidence interval, as well as roughly one standard deviation, in the case of a symmetric distribution. We obtained the result $\phi_{2} = 0.96^{+0.38}_{-0.25}$.

The fitting procedure is influenced by several systematic uncertainties due to: 
\begin{enumerate}[(i)]
 \item The correlation among $\Delta F_2^{(e)}(M)$ values for consecutive $M^2$s. This is in fact taken into account by the resampling method. However, since this treatment is implicit, we employed also  the sparse binning (SB) method\footnote{\mbox{N. Davis} and \mbox{F. K. Diakonos}, in preparation.} as an additional estimator of the systematic error, using in the PF only $\Delta F_2^{(e)}(M)$ values for $M$s that are widely apart in order to avoid strong bin correlations. In this case we used only one sample (the original data set) in the calculations. Changing the distance $\delta M$ between the subsequent $M$s of the $\Delta F_2^{(e)}(M)$ values used in the fit, one obtains a set of $\phi_2$ values which leads to a mean value $\phi_{2,SB} = 0.87$ and a standard deviation $\delta \phi_{2,SB}= 0.08$. We used spacings of $\delta M=2 \ldots 8$.
 \item The $M^2_{min}$-value. We checked the sensitivity of the fitted value of $\phi_2$ to $M^2_{min}$. Similarly as in the SB case, we considered only the original data set and we obtained a set of $\phi_2$ values (one for each $M^2_{min}$, from $6000$ to $12000$ in steps of $1000$) leading to a mean value $\phi_{2,M^2} = 0.82$ and a standard deviation $\delta \phi_{2,M^2} = 0.14$.
 \item The $q_{inv}$-cutoff value. We performed the intermittency analysis changing the minimum of $q_{inv}$ used in the selection of proton pairs in the range $[10,30]$ MeV/c. The corresponding change in the median value of $\phi_2$ was less than $8\%$ indicating that the obtained intermittency result remains practically unaffected by this cut.
\item The proton purity level. Due to insufficient statistics it was not possible to increase the purity of the proton tracks used in the intermittency analysis of the ``Si"+Si system without increasing significantly the errors of the correlator $\Delta F_2^{(e)}(M)$. Thus in our analysis we optimized the choice of the purity level ($80 \%$) taking into account the two competing factors (statistics vs. purity).
\end{enumerate}
According to the preceding discussion the total systematic error, using an euclidean sum of all contributions, was estimated to be $\delta \phi_{2,sys} = 0.16$. For the performed PFs the $\chi^2$ per degree of freedom was below 1.0 (range $[0.09,0.51]$ ) owing to the correlations between the $\Delta F_2(M)$ values of successive M. 

\section{Summary and conclusions}
\label{sect5}

In summary, NA49 performed a search for critical fluctuations employing an intermittency analysis in central ``C''+C, ``Si''+Si and Pb+Pb collisions using second scaled factorial moments of the proton density in transverse momentum spa\-ce. Our analysis demonstrates the presence of non-Poisson\-ian fluctuations in the ``Si''+Si freeze-out state at $\sqrt{s_{NN}} = {17.3\, \text{GeV}}$ ($T \approx 162$~MeV, $\mu_B \approx 260$ MeV) \cite{Becattini2006b}, consistent with a power-law behaviour, as expect\-ed for the emergence of self-similar structured fluctuations \cite{Vicsek} characteristic of the approach to the critical point. No traces of critical correlations were found in the freeze-out states of ``C" + C ($T \approx 166$~MeV, $\mu_B \approx 262$ MeV) and Pb + Pb interactions ($T \approx 155$~MeV, $\mu_B \approx 240$ MeV) \cite{Becattini2006b} at the same collision energy. The power-law exponent
\[
 \phi_{2} = 0.96^{+0.38}_{-0.25} \text{ (stat.)} \pm 0.16 \text{ (syst.)}
\]
for the ``Si"+Si system approaches in size the QCD prediction ($5/6$). An analogous intermittency effect was found recently \cite{NA49_PRC} in central ``Si''+Si collisions at $\sqrt{s_{NN}}$ $= {17.3\, \text{GeV}}$ for $\pi^+ \pi^-$ pairs with invariant mass close to twice the pion mass ($\sigma$-field configurations). Although the freeze-out states of ``C''+C and Pb+Pb are close to that of ``Si''+Si, the critical fluctuation pattern probably cannot develop in ``C''+C due to the small size~\cite{Berdnikov} and the diluteness of the system and may be erased in Pb+Pb during the longer evolution of the hadron phase. The large statistical errors in ``Si''+Si, do not allow a conclusive statement concerning the location of the CP. Nevertheless, the presented intermittency results favour the neighbourhood of the ``Si''+Si freeze-out state for a further detailed search for the CP. Such a program is currently pursued by the NA61 experiment studying A+A collisions with small and intermediate size nuclei.

The  Beam  Energy  Scan  (BES)  program  at  RHIC  \mbox{\cite{STAR, RHIC2, whitepaper}} covers the region $205$ MeV $<~\mu_B~<$ $420$ MeV of the phase diagram  and searches for the CP by studying higher moments of the net-proton multiplicity distribution. The expected signal of the CP is a maximum of the kurtosis times the variance ($K \sigma^2$) of the net-proton multiplicity distribution (a global observable). Our approach here is different since we search for power-law fluctuations of the proton density (a local observable) with a predicted exponent $\phi_2 \simeq \frac{5}{6}$ as a signature of the CP. Certainly, within both approaches, precision measurements are needed in order to reach conclusive results. \\

\begin{small}
\noindent\textbf{Acknowledgements:}\\ This work was supported by the US Department of Energy Grant DE-FG03-97ER41020/A000, the Bundesministerium fur Bildung und Forschung (06F 137), Germany, the German Research Foundation (grant GA\,1480/2-2), the National Science Centre, Poland (grants \mbox{DEC-2011/03/B/ST2/02617} and \mbox{2014/14/E/ST2/00018}), the Hungarian  Scientific  Research  Foundation  (Grants  OTKA 68506, 71989, A08-77719 and A08-77815), the Bolyai Research Grant, the Bulgarian National Science Fund (Ph-09/05), the Croatian Ministry of Science, Education and Sport (Project 098-0982887-2878), Stichting FOM, the Netherlands and the ``Demokritos'' National Center for Scientific Research, Greece.
\end{small}








\end{document}